\documentclass[10pt,twocolumn]{IEEEtran}

\usepackage{graphicx,cite,epsfig,amssymb,amsmath,subfigure,url,stfloats,latexsym}
\usepackage{siunitx}

\begin{document}
\markboth{IEEE Network, Vol. XX, No. Y, Month 2020}
{He, Yang \& Chen: 6G Cellular Networks \ldots}

\title{\mbox{}\vspace{1.5cm}\\
\textsc{6G Cellular Networks and Connected Autonomous Vehicles } \vspace{1.5cm}}
\author{Jianhua~He,~\IEEEmembership{Senior Member,~IEEE}, Kun~Yang,~\IEEEmembership{Senior Member,~IEEE}, and $\mbox{Hsiao-Hwa~Chen,}${$^{^\dagger}$}~\IEEEmembership{Fellow,~IEEE}
\thanks{Jianhua~He (email: {\tt j.he@essex.ac.uk}) and Kun~Yang (email: {\tt kun.yang@essex.ac.uk}) are with the School of Computer Science and Electronic Engineering, Essex University, UK. Hsiao~Hwa Chen (email: {\tt hshwchen@mail.ncku.edu.tw}) is with the Department of Engineering Science, National Cheng Kung University, Taiwan.}
\thanks{This work was supported in part by European Union's Horizon 2020 research and innovation programme under the Marie Skodowska-Curie grant agreement No 824019, the National Natural Science Foundation of China (U1764263 and 61671186), and Taiwan Ministry of Science \& Technology (109-2221-E-006-175-MY3 and 109-2221-E-006-182-MY3).}
}

\date{\today}
\renewcommand{\baselinestretch}{1.0}
\thispagestyle{empty} \maketitle \thispagestyle{empty}


\begin{abstract}
With 5G mobile communication systems been commercially rolled out, research discussions on next generation mobile systems, i.e., 6G, have started. On the other hand, vehicular technologies are also evolving rapidly, from connected vehicles as coined by V2X (vehicle to everything) to autonomous vehicles to the combination of the two, i.e., the networks of connected autonomous vehicles (CAV). How fast the evolution of these two areas will go head-in-head is of great importance, which is the focus of this paper. Based on a brief overview on the technological evolution of V2X to CAV and 6G key technologies, this paper explores two complementary research directions, namely, 6G for CAVs versus CAVs for 6G. The former investigates how various 6G key enablers, such as THz, cell free communication and artificial intelligence (AI), can be utilized to provide CAV mission-critical services. The latter discusses how CAVs can facilitate effective deployment and operation of 6G systems. This paper attempts to investigate the interactions between the two technologies to spark more research efforts in these areas. 
\end{abstract}

\vspace{0.2in}
\section{INTRODUCTION}\label{sec:introduction}

Mobile communication industry is one of very few industry sectors that have been fast-growing for more than three decades. The upcoming 5G mobile networks promise to further change our modern society and vertical industries with three identified services/use cases: eMBB (enhanced mobile broadband), URLLC (ultra-reliable and low-latency communications), and mMTC (massive machine-type communications) \cite{And14}. With 5G mobile systems being commercially rolled out gradually, research discussions on next generation mobile systems, i.e., 6G, have started \cite{Lat19,Yan19,Zha19,Saa20}. 6G mobile network technologies are driven by the challenging demands of emerging mobile applications such as extended reality, industry 5.0 and digital twins, which go beyond the 5G capacity. Though 6G is still in its conceptualization stage, some leading vendors have released initial drafts of technology-driven key performance indicators (KPIs) for 6G, including 1 Tbps peak data rate, 0.1 ms radio latency, max 1 out million outage reliability, 10 times more energy efficiency, 20 years battery life time, and 100 devices per square meter density. There is a consensus on the potential enabling technologies for 6G: THz communications, integrated spatial-terrestrial networks (ISTN), reconfigurable intelligent surface (RIS) and artificial intelligence (AI) \cite{Lat19,Zha19,Saa20,Let19}. These revolution-natured technologies will drive the evolution of existing technologies, such as advanced channel coding and modulation, very large-scale antenna, spectrum sharing and full duplex, etc.

Connected autonomous vehicles (CAVs) \cite{Abo19} is one of the critical vertical industries in 6G with its various demanding service qualities. There could be two levels of definitions for CAV. Basically CAV can mean autonomous vehicles (AV) that are connected to other vehicles and/or infrastructure. AVs are capable of sensing driving environment and moving safely with little or no human control. In an advanced level CAV also refers to the technologies and applications centered around connected autonomous vehicles that can collaborate with each other and infrastructure to achieve improved road safety and efficiency compared to individual AVs without cooperation. Initially, connected vehicles and autonomous vehicles were developed in parallel, which are widely regarded as two most promising technologies for future transportation systems. However, they both have inherent shortcomings. The combination of connected vehicles and autonomous vehicles, thus giving the rise of so-called CAV, has attracted significant momentums to tackle the transportation challenges. There are many promising CAV applications, such as cooperated platooning, smart intersections, and cooperative perception, which can significantly improve road safety and efficiency, fuel consumption and congestion. To unleash the full potentials of CAV, 6G needs to fulfill the following more stringent KPIs for the connectivity of vehicles, including (1) extremely high reliability: 99.999\%; (2) extremely low latency (0.1ms radio latency); (3) extremely massive instant access anytime and anywhere; (4) extremely high throughput to cope with high volume of data for full automation. These KPIs pose significant challenges, calling for new thinking and new communication technologies that go beyond the current ones, such as Long Term Evolution (LTE) enabled C-V2X \cite{Che17} and 5G New Radio (NR) V2X \cite{Lie20,Nai19}. We will discuss these issues in terms of two visional perspectives as follows.

\begin{figure*}[htb]
\centering		
\includegraphics[width=140mm]{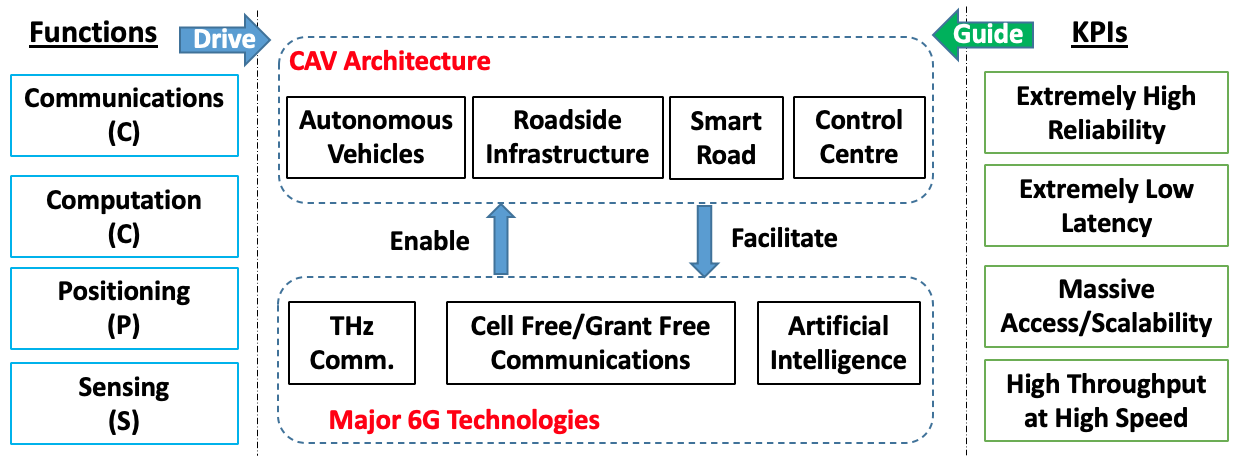}
\caption{Architecture of CAV in 6G Era.}
\label{fig:6G_CAV_design}	 		
\end{figure*}

\begin{itemize}
\item
\textbf{Vision 1}: 6G will be a strong catalyst and enabler for CAV's mission critical services. 5G is targeted mainly for communications. 6G will go beyond communications. Research has been conducted to utilize radio frequency (RF) to carry out object sensing and positioning. With 6G going towards higher spectrum (such as THz) and thus shorter wavelength, a more precise sensing and positioning resolution is theoretically achievable. Higher frequency bands will enhance beamforming directionality and result in increased data throughput. There is also an increasing trend of embedding intelligence (obtained via computation) into communication networks to cope with an increasing complexity of networks and network management. It is our belief that 6G mobile systems facilitate missions of cross functions, combining communications (C), computation (C), positioning (P) and sensing (S) or CCPS for short, in order to satisfy service/application requirements while being deployed at a large scale with cost effectiveness. A joint design of a 6G system multiple cross functions is likely to find a more efficient solution that is beneficial to all CCPS functions, which are also essential to a CAV system to fulfill auto-driving.  
\item
\textbf{Vision 2}: CAVs will facilitate service provisioning and operation of 6G. Roads are an integral part of a modern city, in the same way as vehicles to a family or a society. Stationary roads and mobile vehicles form an important part of our base infrastructure. A CAV system that is composed of smart roads, road side units (RSUs), and vehicles can provide considerable resources, physical space and services for communication, computing and intelligence. The unique CAV features such as controlled mobility and ease of deployment can strongly support the CCPS functions of 6G systems through infrastructure extension, monitoring and maintaining 6G networks, to achieve the 6G goals of ubiquitous wireless intelligence and minimize network operation costs. With expected tremendous investments on the 6G and CAV infrastructures, the 6G communications and CAV systems could be jointly designed, planned and operated with a much better reuse of system resources and services.
\end{itemize}

This paper aims to provide insightful discussions on these two new research fields by bringing them together. This is one step further than the latest ongoing research on 5G NR V2X and IEEE 802.11bd. Moreover, this paper endeavors to identify potential research directions of applying 6G to CAVs and vice versa. The relationship between 6G and CAVs is illustrated in Figure 1. As discussed above, the CAV KPIs and CCPS functions guide and drive the design of a CAV system, which consists of autonomous vehicles, RSU, smart road and also a control centre. CCPS functions and thus the CAV architecture is enabled by 6G technologies such as THz, cell free and AI, etc. In return, CAV infrastructure also facilitates the implementation and deployment of 6G in real life. Of course, the existing 4G and emerging 5G technologies will also carry on to support CAVs. 

\begin{figure*}[htb]
\centering		
\includegraphics[width=160mm]{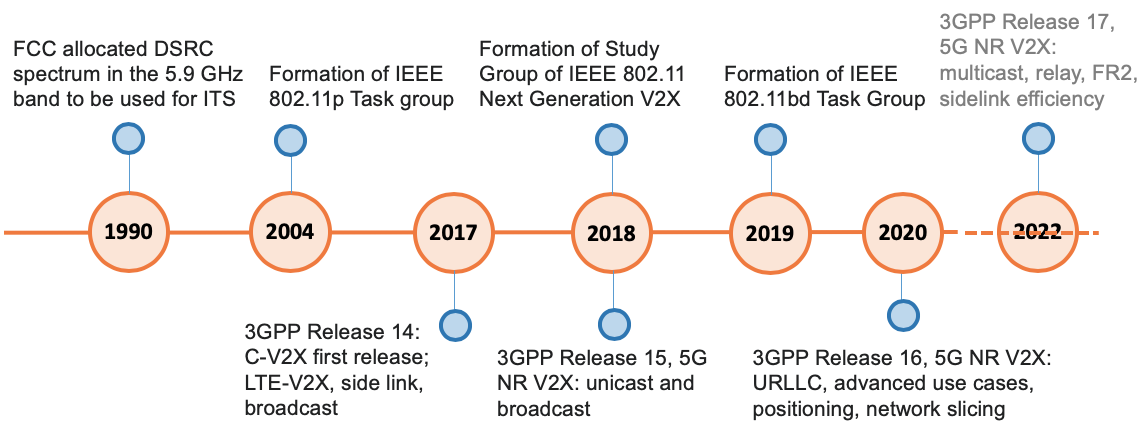}
\caption{A roadmap of V2X technological evolution, where 3GPP Release 17 is in its planning stage and expected to be completed in 2021-2022.}
\label{fig:v2x_evolution}	 		
\end{figure*}

\vspace{0.2in}
\section{EVOLUTION OF V2X AND TO CAV}
\subsection{Evolution of V2X}
Wireless connectivity is critical to CAV applications. Originated from the early research on vehicular ad-hoc networks (VANETs) in the late 1990, vehicle to everything (V2X) has been widely studied and standardized as the cornerstone of connected vehicles. V2X is an umbrella of various communications technologies, covering vehicle to vehicle (V2V), vehicle to infrastructure (V2I), vehicle to pedestrian (V2P), and vehicle to network (V2N). The enabling V2X technology started with dedicated short range communications (DSRC) with spectrum in the 5.9 GHz band allocated by when US Federal Communication Committee (FCC) and Europe ETSI for intelligent transportation systems (ITS). DSRC can be used to support many ITS applications such as toll collection, vehicle safety and in-vehicle entertainment etc. The technology underpinning DSRC is IEEE 802.11p, which is an amendment to the IEEE 802.11 standard to add wireless access in vehicular environments with its particular focus on the PHY and MAC layers. Over the years, various V2X configurations, both in academia and in industries, have been developed, including ITS-G5 in Europe, and they have been based predominantly on IEEE 802.11p.

However, IEEE 802.11p suffers several limitations due to the primary use of random channel access, which include 1) lack of QoS guarantee, 2) unbounded delay at physical channels, 3) short-lived connectivity between vehicles and infrastructure, 4) a need to deploy IEEE 802.11p infrastructure at a large scale, etc. To overcome these limitations, the 3rd generation partnership project (3GPP) proposed cellular based V2X, namely, C-V2X, which utilizes existing cellular communication infrastructure for V2X \cite{Che17}. Unlike IEEE 802.11p, cellular technologies have inherent QoS mechanisms and are well-known for the merit of mobility management. C-V2X started with LTE, which is already commercially deployed widely and used in our everyday life as known as 4G. 

Support of V2X services by 3GPP was specified in the Release 14 standards, which can be provided over PC5 interface by sidelink transmissions. The V2X transmissions are controlled by centralized scheduling via eNodeBs (LTE base station) or gNBs (5G new radio base station) or by distributed scheduling by vehicles. Centralized broadcast scheduling has a high broadcast reliability, but signaling overheads are very high due to vehicle position update and resource allocation. Distributed scheduling with autonomous resource selection has a higher scalability. The sidelink transmission for V2X was enhanced in 3GPP Release 15 with some new functions such as transmission diversity, carrier aggregation and higher quadrature amplitude modulation. The 3GPP Release 16 specifies 5G NR based V2X support with a few more new functions, including support for unicast and multicast, which are particular useful for CAV applications, enhanced channel sensing, resource selection and QoS management schemes, and congestion control \cite{Lie20,Nai19}. 3GPP Release 17 is in its planning stage, which is expected to be completed in 2021, with further enhancement on the sidelink efficiency, URLLC and positioning, and use of relay and frequency range 2 (FR2) over 6 GHz.

While C-V2X is developing rapidly in recent years, IEEE is also catching up. A new Study Group called the IEEE 802.11 Next Generation V2X was formed in March 2018, which led to the formation of IEEE Task Group 802.11bd (TGbd) in January 2019. The ultimate goal of 802.11bd is similar to that of 5G NR-V2X in that they both aim to reduce latency, to improve throughput, and to provide a higher reliability (e.g., 99.999\%). However, their design methodologies are different, namely, 802.11bd requires backward compatibility on the same physical channel, whereas NR-V2X allows communication with its predecessor (LTE-V2X) using a different radio channel \cite{Nai19}. While the benefit of 802.11bd methodology is obvious, it also introduces some huge challenges on its design and performance. A timeline on the evolution of the V2X technologies is illustrated in Figure 2.

\begin{table*}[htb]
\centering 
\caption{Comparison of CAV sensors (LOS: line of sight).}
\label{tab:cav_sensors} 
{\begin{tabular}{ p{1.3cm}| p{3.5cm} | p{3.5cm} | p{1.0cm} | p{1.0cm}| p{1.0cm}| p{1.6cm}  }
\hline
\textbf{Sensors}    & \textbf{Strength} &   \textbf{Shortcomings}     & \textbf{Distance (m)}   
& \textbf{Data rate (Mbps)}    & \textbf{No. of sensors}  & \textbf{Computation} \\					
\hline
Camera	  &  Resolution, rich features, \newline low cost, long range	& No depth, poor weather, LOS detection	& 250	&100s	&10	 & High  \\
\hline
Ultrasound	&Reliable detection	&Short distance, low resolution	&10	 &0.01	&10	 &Very low \\
\hline
Radar	&Resilience, depth and speed	&Low resolution, no height detection	 &300	&10	 &5	 &Low \\
\hline
Lidar	&3D detection, long range, resilience	 &LOS, affected by poor weather	  &250	&10s	&3	&Medium \\
\hline
V2X	       &NLOS, long range, resilience	&Rely on input from other vehicles	&1000	&10	  &1	 &Very low  \\
\hline
\end{tabular}}
\end{table*}

\subsection{Evolution to CAV}
Another line of CAV development is the increased level of automation in vehicles. Starting from basic advanced driving systems (ADS), such as forward collision warning and automatic electronic braking, impressive CAV milestones have been achieved with robotic taxi services offered in the United States and China. The robotic taxis can drive autonomously under certain conditions, which is at Level 4 of the Automated Driving standard set by the Society of Automotive Engineers (SAE) \cite{Qay20}. Many traditional car manufacturers and IT companies, such as Audi, Mercedes Benz and Google, are now striving to achieve the full automation (Level 5 of the SAE automated driving standard). 

According to Mobileye, the largest ADS company, there are three technological pillars for autonomous driving, namely sensing, driving and mapping. The task of sensing is to build environment model with 360 degrees awareness, for example, detection of obstacles and road signs. Various sensors have been utilized in autonomous vehicles, including cameras, ultrasound, lidar, short range and long range radars. All these sensors have their own strengths and shortcomings. For example, cameras have high resolution, long range detection, and capability of recognizing road signs and traffic lights, but have problems when working in poor light and weather conditions. Radar has the benefits of detecting objects with distance and speed, being robust against the poor light and weather conditions, but has shortcomings of low resolution and is not able to recognize the shape and height of the detected objects. Lidar has wide field of vision and accurate distance estimation, but has poor resolution and is affected by bad weather conditions. 

To enable safe and reliable driving under various challenging road, weather and light conditions, different strategies with a mixture of sensors have been chosen for vehicles. For example, radar, lidar in addition to cameras are used in the latest Waymo autonomous driving platform \footnote{https://waymo.com/tech/}. Tesla autonomous vehicles uses cameras, radar and ultrasonic, but not lidar.  Fusion of multiple local sensors could improve sensing and safety, but they are still limited by line of sight (LOS) detection and the detection performance degrades with object distance. Even though autonomous driving companies are investing heavily on autonomous vehicles with increasingly powerful and more sensors, the full driving automation could still be out of reach due to many factors, such as limitations of machine learning algorithms and sensors, challenging driving conditions and road emergencies, lack of redundancy on sensor safety and infrastructure support. It turns out that connected vehicle is an excellent complementary technology to autonomous vehicles and is widely regarded as an integral part of the fully autonomous vehicles. CAV as a combination of both can address most of the aforementioned challenges faced by autonomous vehicles alone on the road to full automation. Furthermore, apart from enhancing driving safety by cooperative perception and cooperative driving, CAV can enable many new road efficiency applications, such as cooperative platooning and remote driving \cite{Abo19}. CAV has gain a huge momentum from automotive and telecommunication industries, academia and public authorities in the last several years. A brief summary of sensors and connectivity used in CAV and their features is presented in Table.1.

\vspace{0.2in}
\section{6G SUPPORT FOR CAV}

\begin{figure*}[htb]
\centering		
\includegraphics[width=142mm]{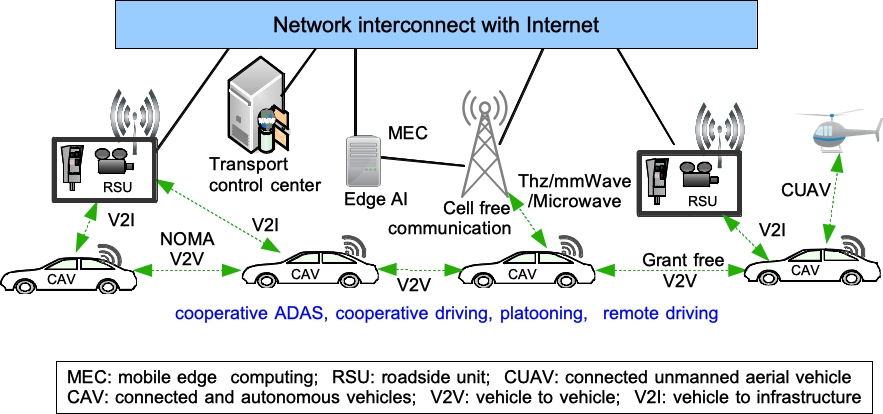}
\caption{CAVs in 6G Communications.}
\label{fig:6G_for_CAV}	 		
\end{figure*}

A typical CAV system, as shown in Figure \ref{fig:6G_for_CAV}, may include key components of CAVs on roads, RSUs (equipped with communication, computing and traffic control devices), smart roads with intelligent materials and sensors, and transport control center. The RSUs will play a critical role on collaborative mobility and computing. In the meanwhile, unmanned aerial vehicles (UAVs) are applied in many scenarios to supplement on-road vehicles. Connected unmanned aerial vehicles (CUAVs) are also regarded as a part of a CAV system. The new functions such as CCPS that could be offered by 6G systems can bring in significant benefits to CAV from both connectivity and computing aspects, as also assisted by key technologies existing in 5G networks (such as mmWave, massive MIMO, network virtualization function or NFV, and software defined networks or SDNs).

\subsection{Key 6G Enabling Technologies}

6G is envisioned to provide unprecedented capacity, sub-millisecond latency and ubiquitous coverage \cite{Lat19}. It not only targets at Tbps data rates, but also inherently supports a wide range of novel scenarios and applications that combine agility, reliability, ultra-low latency, and energy efficiency. There are already worldwide 6G research initiatives and programs, in which various technologies and roadmaps have been proposed. In July 2018 ITU formed a research group on 2030 network technologies. South Korea wireless telecommunications operator SK Telecom proposed a 6G technical roadmap of THz, cell free networks, and airborne wireless platforms in 2018. It agreed with Samsung to work on joint 6G evolution technologies in 2019. China started 6G research initiatives to define Beyond 5G vision and requirements in 2018. Huawei also announced the research study of 6G technologies in 2019. In 2018, FCC envisioned 6G as featuring THz networks with multiple simultaneous beams of data transfer, requiring unprecedented network densification.  

University of Oulu started the Finnish 6G Flagship program and organized the first 6G wireless summit in 2019. They published a 6G white paper \cite{Lat19}, which identified key drivers, research requirements and challenges. W. Saad \textit{et al.} \cite{Saa20} discussed about the 6G vision and technologies in a much broader sense. According to the reported 6G programs and technical roadmaps, THz, reconfigurable intelligent surface (RIS), ISTN, AI and distributed computing have been tipped to be the most promising candidate technologies for 6G. B. Letaief \textit{et al.} \cite{Let19} painted a 6G roadmap with particular emphasis on the empowerment of AI on 6G. Figure 3 illustrates how various 6G technologies can be integrated into CAV architecture and support CAVs. Some of these candidate technologies relevant to CAVs will be discussed next.

\subsection{THz for CAVs}
Exploitation of the spectrum in the THz regime is vital to achieve the 6G KPI of 1 Tbps data rate \cite{Lat19,Zha19}. THz band usually refers to the frequency band in the range of 100 GHz to 1 THz with the corresponding wavelength of  \SI{30}{\micro\metre} to 3 mm. The rich frequency resource available at the THz band can provide a large bandwidth (up to 10s GHz) and enable high speed links needed for the data intensive communications between CAVs and between CAVs and infrastructure. Many advanced CAV applications can be supported, including cooperative perception with raw sensor data, mobile edge intelligence and remote driving. In addition to the communication benefits, 6G systems operating on the THz band hold great potentials on positioning, sensing and 3D imaging \cite{Lat19}. As THz signals have short wavelength, the antenna size and separation distance can be reduced radically, which allows a large antenna array to be installed in mobile devices and base stations. For example, more than 200 antennas can be fit into an area of 1 square cm at 300 GHz. Narrow beams can be generated with a large antenna array, which enables precision positioning for mobile devices with an outdoor localization error of less than 1 meter. The high carrier frequencies of THz can also enable radio frequency based sensing, providing accurate position and object detection. The localization and sensing are highly appealing to CAVs. Traditional GPS satellite based positioning method suffers low localization precision and limited performance in urban environments. On the other hand, high definition mapping is not highly responsive to the changes of roads such as road construction and maintenance. 6G networks with THz can provide a low cost, precision and reliable alternative to localization and sensing for CAVs, which is extremely important to realize the full automation level. However, there will be major challenges on designing THz antenna/transceiver, increasing communication range, and dealing with high CAV mobility.

\subsection{Grant Free and Non-orthogonal Multiple Access (NOMA)}
6G systems are expected to meet the unprecedented requirement of supporting a massive number of IoT devices while attaining ultra-reliability and low latency. A key enabling and candidate technology for 6G to meet the requirement is NOMA, which was proposed for 5G but yet adopted by 3GPP. It allows multiple users utilize non-orthogonal resources concurrently for both random access and multiplexing. NOMA can be achieved in power domain, code domain and pattern domain. NOMA has been successful applied to grant free access approaches \cite{Cir19}\cite{Guo19}, which support massive connectivity and achieve performance close to scheduled access schemes. Compared to orthogonal multiple access, NOMA also has a superior spectrum efficiency. 

NOMA is complementary to other 6G candidate technologies such as THz communication and can be used for both V2V and V2I communications. It can be adopted in an enhancement to the 5G V2X technologies and bring great benefits to CAV applications. CAVs usually operate under high vehicle density and mobility, long communication range and heavy traffic conditions. NOMA could contribute with its capability of supporting massive connectivity and superior spectrum efficiency. For example, sparse code multiple access (SCMA) can be applied to data resource reservation in the random access channels and data packet transmissions over data channels in the 3GPP V2V protocols \cite{Cir19}. Research works have been reported on the application of NOMA to V2X, which are focused mainly on the V2I links and centralized resource allocation \cite{Di17}. Technical challenges include the design of practical and efficient NOMA based grant free access and multiplexing schemes for CAV applications, especially in distributed V2V networks scenarios.

\subsection{Cell Free Communications} 
To satisfy the need of higher rates in 6G networks, exploitation of mmWave and THz frequency bands will be necessary. The cells operating in these frequency bands are small with a radius of a few tens meters and the communication is subjected to large path loss and signal block by obstacles. mmWave, THz and microwave communication technologies will be integrated with co-existing multiple scales of cells, leading to frequent handovers in such multiple scale networks. It is more challenging for CAVs due to the strict requirements of CAVs on communication and safety. 

A promising approach to tackle the mobility problem for the 6G multi-scale networks is cell free communication technology \cite{Ngo17}. In a cell free communication system, a large number of access points (each with one or a few number of antennas) are deployed and they cooperate via a backhaul network and a central processing station to serve all mobile users distributed over a wide area. In such systems, there are no cells or cell boundaries. The mobile users move seamlessly within the heterogeneous networks and receive the optimized service qualities. CAV applications can benefit by the cooperative communication technologies to avoid the aforementioned handover problems. The cell free networks also offer potential high precision localization and sensing services from THz technology or additional sensors. The integrated cell free communication, localization and sensing will boost the performance of CAV applications with improved cooperative perception and positioning. Major technical challenges include coordination of transmissions from the access points with high vehicle mobility and radio resource management by the central processing unit.

\subsection{Artificial Intelligence (AI) and Edge Intelligence (EI)}
Driven by recent breakthrough on deep learning and its successful applications in many areas such as computer vision and natural language processing, there have been growing interests in the application of AI in mobile networks. It is envisioned that AI will play a vital role in 6G and be applied to many applications, such communications and networking, resource management, network control and automation \cite{Lat19,Let19}. In addition, 6G is expected to provide edge intelligence (EI) services through mobile edge computing (MEC) to IoT devices. CAVs rely heavily on the use of AI, for example, in the environment perception for accident avoidance, high-definition map for navigation and autonomous driving decision making. The heavy computing loads of autonomous driving related applications can be offloaded to the MEC stations via high speed 6G links. For example, CAVs can forward sensor data to MECs, which process the data and aggregate sensing outcomes from other CAVs, then make real-time driving decisions for the CAVs or simply return the sensing outcomes such as obstacles or hazards on the roads. Distributed AI and federated learning can also be supported with 6G MEC for CAV cooperative learning and model training to preserve user privacy.

\begin{figure}[htb]
\centering		
\includegraphics[width=95mm]{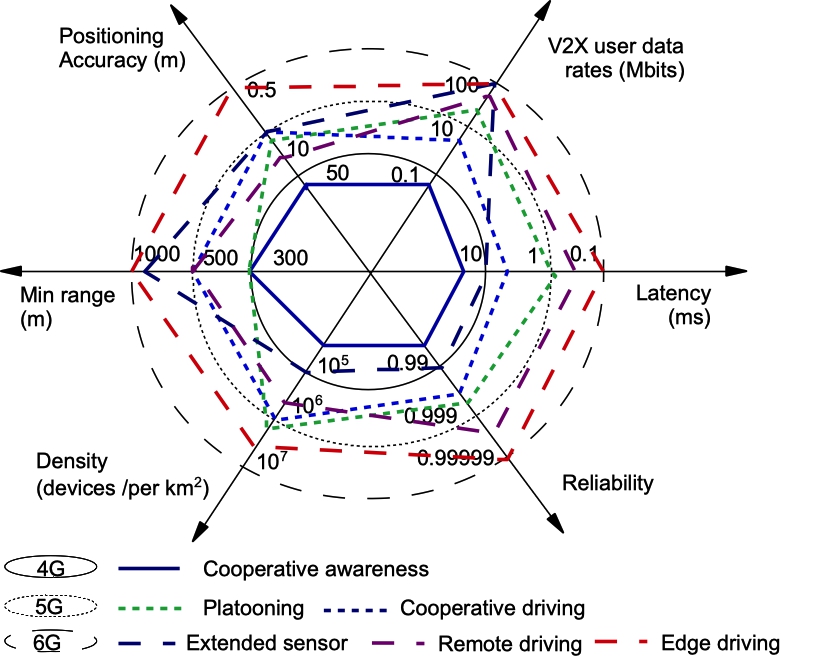}
\caption{Representative CAV applications enabled by 4G, 5G and 6G cellular networks, which may diverse various QoS performance.}
\label{fig:6G_for_CAV_web}	 		
\end{figure}

Apart from the above discussed technologies, many others, such as QoS control, network slicing and block-chain, could also be vital to ensure the strict requirements of CAV mission critical services. Figure~\ref{fig:6G_for_CAV_web} illustrates representative CAV applications that can be enabled by 4G, 5G and 6G cellular networks, which require diverse QoS performance metrics. 6G networks can support advanced uses cases, such as remote driving, edge driving and digital twins of CAVs. Remote driving refers to the CAV applications, where the vehicles are controlled remotely by human or machine drivers. Edge driving means autonomous vehicles are controlled by the machine drivers from the nearby CAVs or edge computing stations. With the advanced 6G support, it is also possible to create digital twins of CAVs, which will collect CAV sensor data, build predictive models for the CAVs, provide edge computing services, control and support the CAVs.

\vspace{0.2in}
\section{CAV ENHANCEMENTS IN 6G NETWORKS}

\begin{figure}[htb]
\centering		
\includegraphics[width=90mm]{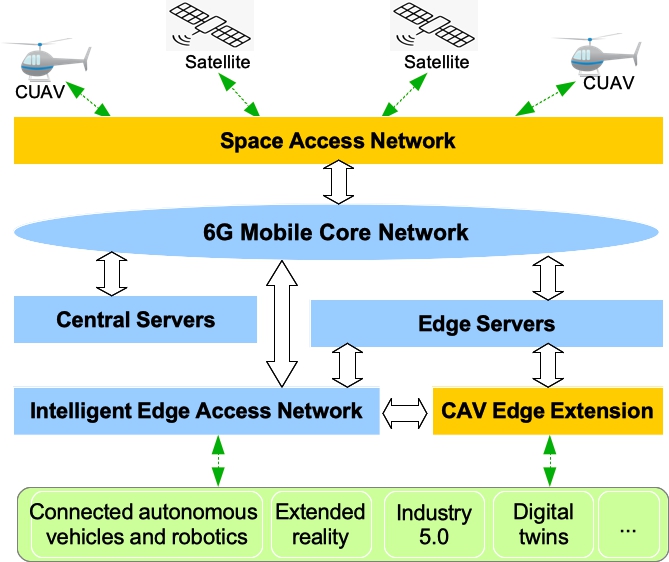}
\caption{CAVs in 6G network architecture.}
\label{fig:CAV_for_6G}	 		
\end{figure}

While 6G will be a strong catalyst and enabler for CAVs, CAV in return can provide a strong support to 6G in the delivery of communication, networking, computing and management services. It will facilitate the 6G systems to achieve the visions of ubiquitous wireless intelligence. In this paper we envision a 6G network architecture as shown in Figure \ref{fig:CAV_for_6G}, which includes mobile core network, space access network, edge clouds and servers, edge access networks and eventually end devices. The mobile core network plays a similar role as the gNBs of 5G networks, to manage mobility, network and connections to the Internet. The space access network provides connectivity to the users via satellites and UAVs. Central and edge servers provide networking and computing services with features of network slicing and NVF. Edge access network connects mobile users and wireless end devices to mobile core network and edge servers. Apart from the support of mmWave and THz communications such as beam alignment and connection management with CAV sensing, the CAV system can enhance the 6G networks from the edge and space access aspects.

\subsection{Extension of 6G Communication Infrastructure}

Dense 6G base stations will be deployed, which demand tremendous investment and are not flexible to respond to fast spatial and temporal changes of cellular network traffic patterns. In addition, under emergency situations such as natural disasters the communication infrastructure may be damaged or not available. The surface CAVs and CUAVs provide an excellent mobile platform to extend the 6G communication infrastructure, which can be converted to CAV mobile base stations (CBS) and deployed to the locations as needed to provide a wider coverage. CBS provides a flexible and economic solution to augment the fixed 6G communication infrastructure. The CBSs can be self-organizing and cooperate with each other, deciding where and when to provide communication services adaptively in response to changing demands. While it is possible that some CBSs are dedicated to providing communication services, most of them are expected to have major duty of providing mobility services.
While it is possible for the CAVs to provide access service in moving, they are assumed to provide the service when they are stationary and play a supportive role by offloading a small part of traffic instead of a major part of traffic from the legacy base stations. The supporting CAVs will establish connections to the base stations for wireless backhaul purpose.
Interestingly, road traffic and wireless network traffic generate patterns which are suitable for the reuse of the CAVs on providing communication services. For example, at night time there is less traffic on roads but more wireless network traffic, thus fewer CAVs are needed for transportation and more CAVs can be used as CBS. The utilization of the CAVs can be largely improved and the 6G network operational costs can be reduced significantly. 

\subsection{Mobile Vehicle Edge Computing}

A key feature of 6G is the provisioning of intelligent services and applications. A cloud computing continuum (3C) consisting of remote clouds and edge clouds will be used to deal with the edge computing applications with diverse computing performance requirements. Due to the mobility of the edge computing service users, computing and data migration will be challenging under the 3C paradigm. With an increasing computing power, which is needed for high level automation of CAVs, CAVs can support 6G networks on providing mobile computing services. The availability of storage space provides an opportunity for CAVs to be equipped with more resources for larger scale vehicle edge computing (VEC) services. They can be deployed according to the demands of 6G network operators or individuals, supporting the other CAVs or other types of mobile users. VEC with features such as autonomous mobility and ease of deployment represents a flexible and great enhancement to 6G MEC.

\subsection{Network Performance Monitoring and Sensing for 6G}

6G networks will have sophisticated structure and require complex management and control. There is a trend of increasing network automation and resilience from the telecommunication equipment vendors and network operators \cite{Hua19}. CAV can contribute to the automation of 6G networks. Performance measurement and monitoring are critical to achieve intelligence control and automation, for which the CAVs and the smart roads with intelligent sensors can contribute with real time update on service quality and surveillance of network infrastructure. The CAVs and smart roads can support the construction and operation of the 6G RIS based smart communication environment. They can be deployed on demand to measure network service quality, act as temporal base stations to enhance 6G communications, detect and even repair problematic outdoor 6G network components. With increasing penetration of CAVs they can provide economic alternative to the slow and costly manual network monitoring and enhance 6G automation.

\vspace{0.2in}
\section{CONCLUSIONS}
	
This paper brought together two promising research directions, i.e., CAVs and 6G networks, and discussed their potential interactions and mutual support. After drawing out a roadmap of the technical evolution of V2X to CAV and that of 6G key technologies, this paper explored two complementary directions of future researches, namely, 6G for CAVs and CAVs for 6G. Discussions have been made to show how various 6G key enablers such as THz, cell free communications and edge intelligence can be utilized to enable CAV's mission-critical services. Proposals are also made to illustrate how CAVs can be employed for a more effective and efficient deployment and operation of 6G systems. This is our belief that the intersection of CAV systems and 6G networks will bring in significant innovations and momentum to both areas. A joint design of both may be an effective way forward and such a consideration shall be taken as much as possible from the early design stage of each, aiming to achieving an optimized integral system that benefits both sectors.

\vfill

\end{document}